\documentclass{aip-cp}

\usepackage[numbers]{natbib}
\usepackage{xspace}
\usepackage{aas_macros}
\usepackage{url}

\newcommand{\pks}{PKS\,2155$-$304\xspace}
\newcommand{\pg}{PG\,1553$+$113\xspace}
\newcommand{\hess}{H.E.S.S.\xspace}
\newcommand{\fermi}{\textit{Fermi}-LAT\xspace}
\newcommand{\phasetwo}{\hess~II\xspace}

\newcommand\arcmin{\mbox{$^\prime$}}%
\newcommand\arcsec{\mbox{$^{\prime\prime}$}}%
\newcommand\degr{\mbox{$^\circ$}}%
  
\begin{document}

\title{Gamma-Ray Blazar Spectra with \phasetwo Mono Analysis: The Case of \pks and \pg}

\author[aff1]{D.~Zaborov\corref{cor1}}
\author[aff2]{A.M.~Taylor}
\author[aff3]{D.A.~Sanchez}
\author[aff4]{J.-P.~Lenain}
\author[aff2]{C.~Romoli}
\author{\hess~Collaboration}

\affil[aff1]{Laboratoire Leprince-Ringuet, Ecole Polytechnique, CNRS/IN2P3, F-91128 Palaiseau, France}
\affil[aff2]{Dublin Institute for Advanced Studies, 31 Fitzwilliam Place, Dublin 2, Ireland}
\affil[aff3]{Laboratoire d'Annecy-le-Vieux de Physique des Particules, Universit\'{e} Savoie Mont-Blanc, CNRS/IN2P3, F-74941 Annecy-le-Vieux, France}
\affil[aff4]{Sorbonne Universit\'es, UPMC Universit\'e Paris 06, Universit\'e Paris Diderot, Sorbonne Paris Cit\'e, CNRS, Laboratoire de Physique Nucl\'eaire et de Hautes Energies (LPNHE), 4 place Jussieu, F-75252, Paris Cedex 5, France}
\corresp[cor1]{Corresponding author: zaborov@llr.in2p3.fr}

\maketitle

\begin{abstract}
The addition of a 28 m Cherenkov telescope (CT5) to the H.E.S.S. array extended the experiment's sensitivity to lower energies, 
providing new opportunities for studies of Active Galactic Nuclei (AGNs) with soft intrinsic spectra and at high redshifts. 
The high-frequency peaked BL Lac objects PKS 2155-304 ($z = 0.116$) and PG 1553+113 ($0.43 < z \lesssim 0.58$) are among the brightest objects in the gamma-ray sky,
both showing clear signatures of gamma-ray absorption at $E > 100$ GeV interpreted as being due to interactions with the extragalactic background light (EBL).
Multiple observational campaigns of \pks and \pg were conducted
during 2013 using the full H.E.S.S. II array (CT1--5).
To achieve the lowest energy threshold, a monoscopic analysis of the data taken with CT5 was developed
along with an investigation into the systematic uncertainties on the spectral parameters which are derived from this analysis.
The energy spectra were reconstructed down to energies of 80 GeV for PKS 2155-304,
which transits near zenith, and 110 GeV for the more northern PG 1553+113.
The measured spectra, well fitted in both cases by a log-parabola spectral 
model (with a 5$\sigma$ statistical preference for non-zero curvature for \pks and 
4.5$\sigma$ for \pg), were found consistent with spectra derived from contemporaneous 
\fermi data, indicating a sharp break in the observed spectra of both sources at 
$E \approx 100$ GeV.
When corrected for EBL absorption, the intrinsic spectrum of \pks was 
found to show significant curvature. For \pg, however, no significant detection of 
curvature in the intrinsic spectrum could be found within statistical and systematic 
uncertainties.

\end{abstract}

\section{Introduction}
\label{intro}

The very high energy (VHE, $E\gtrsim 100$~GeV) gamma-ray experiment of 
the High Energy Stereoscopic System (\hess) consists of five imaging 
atmospheric Cherenkov telescopes (IACTs) located in the Khomas Highland 
of Namibia ($23\degr 16\arcmin 18\arcsec$ S, $16\degr 30\arcmin 
00\arcsec$ E), 1835~m above sea level. From January 2004 to October 
2012, the array was operated as a four telescope instrument (\hess I).
The telescopes, CT1--4, each have an effective mirror surface area of 107~m$^{2}$ \cite{2006A&A...457..899A}.
In October 2012 a fifth telescope, CT5, placed at the centre of the original square, started 
taking data. This set-up is referred to as \phasetwo.
With its effective mirror surface close to 600~m$^{2}$ and a 
fast, finely pixelated camera \cite{2014NIMPA.761...46B}, CT5 
potentially extends the energy range covered by the array down to 
energies of $\sim 30$~GeV.

In this study, we focus on obtaining high statistic results with observations of the 
high-frequency peaked BL Lac (HBL) objects \pks and \pg. These blazars are among the 
brightest objects in the VHE gamma-ray sky. Furthermore, the spectra of both these
blazars exhibit  signatures of gamma-ray absorption at energies $E \sim 100$~GeV,
due to interactions with the extragalactic background light (EBL).

\pks is located at redshift $z=0.116$ \cite{2013MNRAS.435.1233G}.
The first detection of VHE emission from this object was attained 
in 1996 by the University of Durham Mark 6 Telescope \cite{1999ApJ...513..161C}.
Starting from 2002 the source was 
regularly observed with \hess, with the first detection based on the 
2002 data using just one telescope of \hess I \cite{2005A&A...430..865A}.
After completion of the array, this 
source was detected in stereoscopic mode in 2003 with high significance 
($> 100\,\sigma$) at energies greater than 160 GeV 
\cite{2005A&A...430..865A}. Strong flux variability with multiple 
episodes of extreme flaring activity in the VHE band were reported 
\cite{2007ApJ...664L..71A,2010A&A...520A..83H}. A photon index of 
$3.53 \pm 0.06_{\rm{stat}} \pm 0.10_{\rm{syst}}$ was obtained from 
analysis of observations during a low flux state (2005--2007) above 
200~GeV \cite{2010A&A...520A..83H}. For average and high flux 
states the presence of curvature or a cut-off was favoured from the 
spectral fit analysis carried out \cite{2010A&A...520A..83H}.

The HBL object \pg was first announced as a VHE gamma-ray source by 
\hess \cite{2006A&A...448L..19A} and independently and almost 
simultaneously by MAGIC using observations from 2005 
\cite{2007ApJ...654L.119A}. The \hess~I measurements 
\cite{2008A&A...477..481A} yielded a photon index $\Gamma = 4.5 \pm 
0.3_{\rm{stat}} \pm 0.1_{\rm{syst}}$ above 225~GeV.
The redshift of \pg is constrained by UV observations 
to the range $0.43 < z \lesssim 0.58$ \cite{2010ApJ...720..976D}. 
Assuming that the 
difference in photon indices between the high energy (HE) and VHE regimes is 
imprinted by the attenuation by the extragalactic background light, the 
redshift was constrained to the range $z = 0.49 \pm 0.04$ 
\cite{2015ApJ...802...65A}.

This paper reports on the first observations of \pks and \pg conducted in 2013 
using the \phasetwo instrument (CT5), analysed in monoscopic mode. 
Particular emphasis is placed on the spectral measurements at low 
energies and their connection with the \fermi measurements. Using the
\phasetwo mono and \fermi results, the implications on intrinsic source spectrum are 
considered.

\section{\phasetwo Analysis and Results}
\label{hess2_analysis}

\pks was monitored with \phasetwo regularly in 2013, from April 21 to November 5 (MJD 56403--56601).
\pg was observed with \phasetwo between May 29 and Aug 9, 2013 (MJD 56441--56513).
Most of the observations were taken using the full \phasetwo array.
However this paper is focused on the monoscopic analysis of these data, which provides the lowest achievable energy threshold.

To ensure the quality of the AGN data sets for the \phasetwo mono analysis 
a set of run quality criteria was applied as described in \cite{2015arXiv150906509Z}.
After the quality selection, the \pks and \pg data sets comprise, respectively, 43.7~hr and 16.8 hr of data (live time).
During these observations, the zenith angle of \pks ranged from 7$\degr$ to 60$\degr$, 
with a median value of $16\degr$.
The \pg zenith angle ranged between $33\degr$ and $40\degr$, with a mean value of $35\degr$.
The data sets were processed with the standard \hess analysis software using the Model reconstruction \cite{2009APh....32..231D}
which was recently adapted to work with monoscopic events \cite{2015arXiv150902896H}.
The optimized analysis provides an angular resolution of $\approx$ 0.15$\degr$ (68\% containment radius) at 100 GeV and energy resolution of $\approx$ 25\%.
The background subtraction was performed using the standard algorithms used in \hess \cite{2007A&A...466.1219B}, with only minor adjustments \cite{2015arXiv150906509Z}.
The significance of the excess after background subtraction is determined using the Li \& Ma prescription \cite{1983ApJ...272..317L}.
Spectral measurements are obtained using the forward folding technique \cite{2001A&A...374..895P}.
The \phasetwo mono analysis was applied to all events that include CT5 data (ignoring information from CT1--4).

\begin{figure}
  \centering
\includegraphics[width=0.5\linewidth]{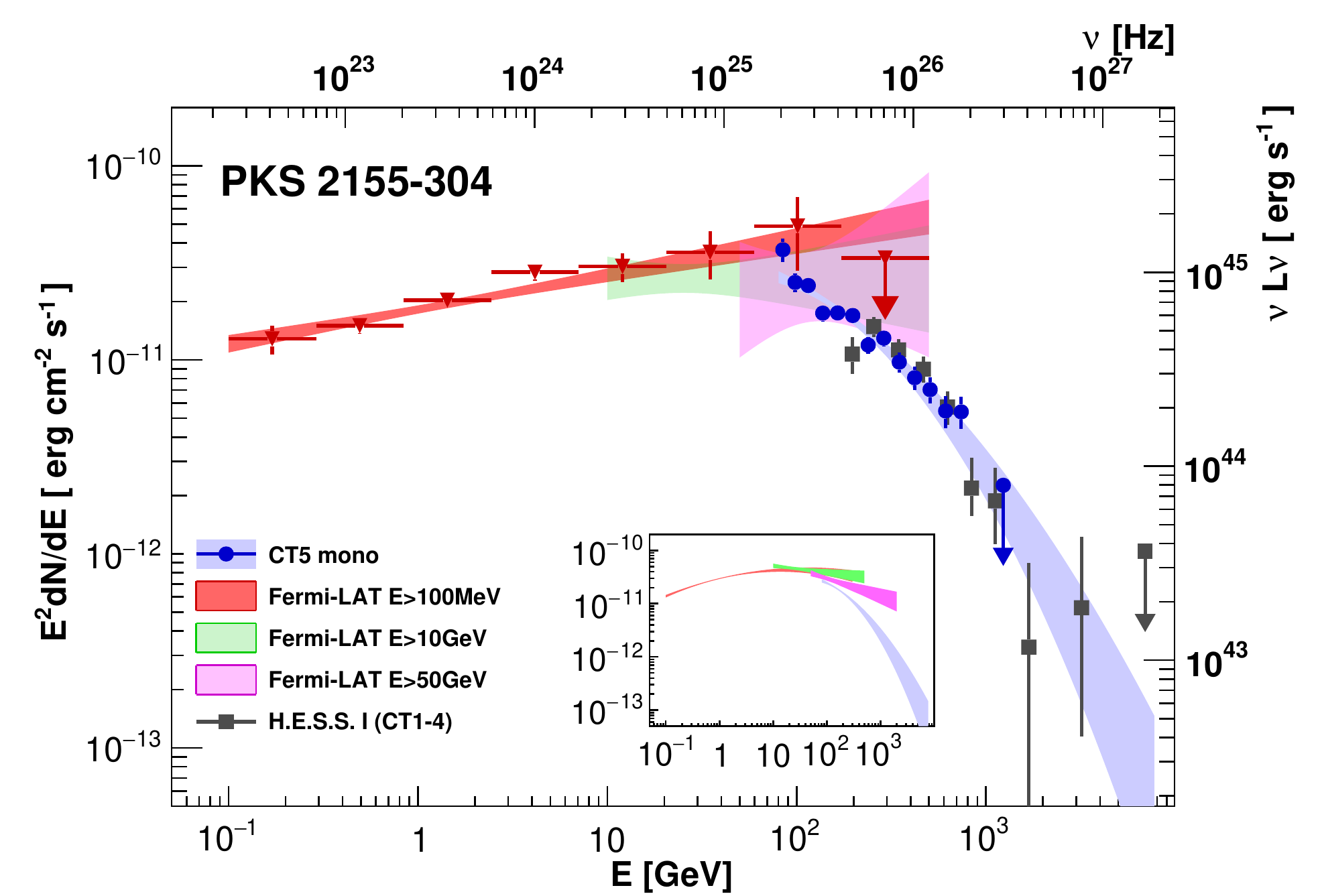}
\includegraphics[width=0.5\linewidth]{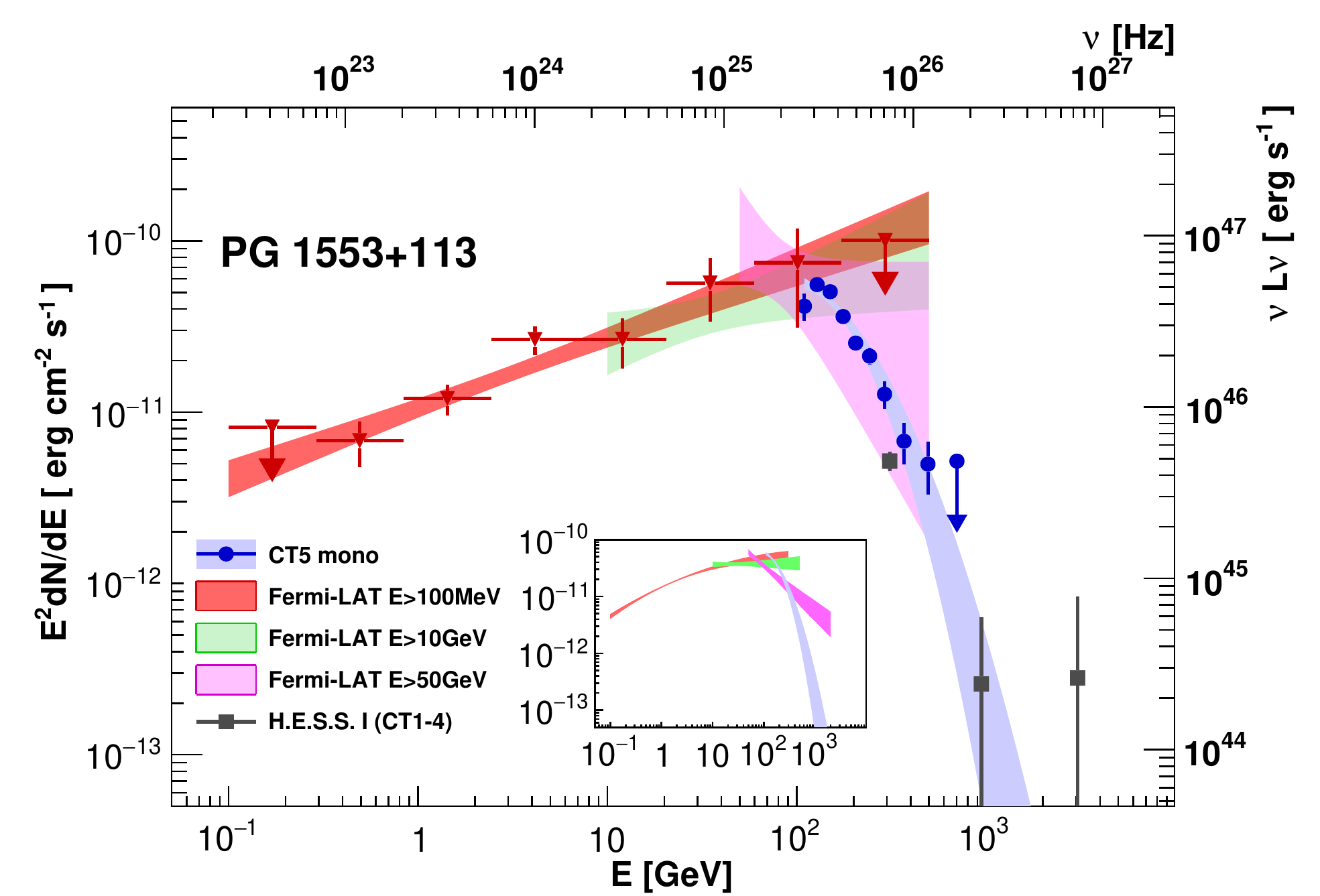}
\caption{The energy spectra of \pks (left panel) and \pg (right panel) obtained from the \phasetwo mono analysis (shown by blue circles with confidence band) in comparison with the contemporaneous \fermi data with an energy threshold of 0.1~GeV (red triangles and confidence band), 10~GeV (green band), and 50~GeV (purple band) and contemporaneous CT1--4 data  (grey squares). In all cases the confidence bands represent the 1~$\sigma$ region. 
The right-hand y-axis shows the equivalent isotropic luminosity (not corrected for beaming or EBL absorption; the assumed redshift of \pg is $z=0.49$).
The insets compare the \hess confidence bands with the \fermi catalogue data (3FGL, 1FHL and 2FHL).}
\label{fig:sed}
\end{figure}

The analysis showed that \pks is detected with a statistical
significance of $\approx 36 \, \sigma$, with $\approx$ 3000 excess events. 
The reconstructed spectrum of \pks is shown in Fig.~\ref{fig:sed}, left. 
A log-parabola model, ${\rm d}N/{\rm d}E = \Phi_{0}\,(E/E_0)^{-\mathbf{\Gamma}-\beta\cdot \log(E/E_0)}$,
better fits the data with respect to a simple power-law model with a log-likelihood ratio of 16 (i.e. a $4\sigma$ preference).
The flux normalisation, above a threshold of 100~GeV, is found to be $\Phi_{0} = (5.30 \pm 0.18_{\rm stat}) \times 10^{-10} \, \rm{cm}^{-2} \, \rm{s}^{-1} \, \rm{TeV}^{-1}$
at a decorrelation\footnote{For the log-parabola model, the decorrelation energy is the energy where the error on the flux is the smallest, i.e. where the confidence band butterfly is the narrowest in the graphical representation.} energy $E_{0} = 151$~GeV,
with a photon index $\Gamma = 2.65 \pm 0.09_{\rm{stat}}$ and a curvature parameter $\beta = 0.22 \pm 0.07_{\rm{stat}}$.
The spectral data points (blue filled circles in Fig.~\ref{fig:sed}) cover the energy range from 
80~GeV to 1.2~TeV (not including upper limits).
A simple power-law fit to the \pks data yields a photon index $\Gamma= 2.93\pm0.04_{\rm{stat}}$.
The observed flux level agrees with the level reported for the 
quiescent state observed with \hess (at $E > 300$~GeV) from observations during 2005--2007 
\cite{2010A&A...520A..83H}.

\pg is detected with a statistical significance of 27~$\sigma$, with $\approx$ 2500 excess events. 
The reconstructed spectrum, with a threshold of 110~GeV, is found to be well fit by a log-parabola
(with a log-likelihood ratio of 20 over the power-law model, see Fig.~\ref{fig:sed}, right), with a photon index $\mathbf{\Gamma} = 2.95 \pm 0.23_{\rm{stat}}$ 
at decorrelation energy $E_{0} = 141$~GeV, curvature parameter $\beta = 1.04 \pm 0.31_{\rm{stat}}$,
and differential flux $\Phi_{0} = (1.48 \pm 0.07_{\rm{stat}}) \times 10^{-9} \, \rm{cm}^{-2} \, \rm{s}^{-1} \, \rm{TeV}^{-1}$ at $E_{0}$.
The spectral data points cover the energy range from 100~GeV to 
500~GeV (not including upper limits). 
These results are in reasonable agreement with the earlier measurements by \hess \cite{2008A&A...477..481A,2015ApJ...802...65A},
MAGIC \cite{2007ApJ...654L.119A,2010A&A...515A..76A,2012ApJ...748...46A} and VERITAS \cite{2015ApJ...799....7A}, 
indicating that the source was in an average state of activity during the 2013 observation campaign.
No significant night-by-night or weekly variability is found in the \phasetwo mono lightcurve.

The systematic uncertainties on the spectral parameters derived in the \phasetwo mono spectral analysis
were estimated through an extensive set of dedicated studies and are summarised in Table~\ref{SystErrorTable}. 
The table gives the flux normalisation uncertainty, 
the photon index uncertainty and the uncertainty on the curvature parameter $\beta$ (for the log-parabola model).
In addition, the energy scale uncertainty is given in the second column.
The energy scale uncertainty implies an additional uncertainty on the flux normalisation which depends on the steepness of the spectrum.
It is also relevant for the determination of the position of spectral features such as the SED maximum or EBL cutoff.
The procedures used here for estimating the systematic uncertainties generally repeat the procedures used for \hess I \cite{2006A&A...457..899A}.
A more detailed description of the systematic uncertainties will be provided in a forthcoming publication.

\begin{table*}[t]
\caption{Estimated systematic uncertainties in the spectral measurements using \phasetwo mono for the analyses presented in this work.}
\label{SystErrorTable}
\centering
\begin{tabular}{c c c c c}
\hline
Object Name & Energy Scale & Flux & Index & Curvature\\
\hline
\pks					& $\pm$ 19\%	& $\pm$ 20\%	& $\pm$ 0.17	& $\pm$ 0.12\\
\pg					& $\pm$ 19\%	& $\pm$ 22\%	& $\pm$ 0.65	& $\pm$ 1.0\\
\hline
\end{tabular}
\end{table*} 

The CT1--4 stereoscopic data collected simultaneously with the CT5 data
have been analysed using the \hess~I version of the Model analysis method 
\cite{2009APh....32..231D}, yielding also significant detections of both sources.
In both cases the spectrum is well fitted by a power-law model.
The resulting forward-folded data points are shown on Fig.~\ref{fig:sed} (grey squares).
The CT1--4 results for \pks were found to be in excellent agreement with the \phasetwo mono results.
Due to the limited statistics and relatively high energy threshold of the CT1--4 analysis,
the CT1--4 results for \pg are represented on Fig.~\ref{fig:sed} by 3 data points only.
Taking into consideration the systematic uncertainties on the energy scale and flux normalization,
the CT1--4 data were found to be in satisfactory agreement with the CT5 results.

\subsection{HE Gamma-Rays Observed by \fermi}
\label{fermi_data}

The \fermi detects gamma-ray photons above an energy of 100~MeV. Data 
taken contemporaneously with the \phasetwo observations were analysed 
with the publicly available ScienceTools {\tt v10r0p5}\footnote{See 
\url{http://fermi.gsfc.nasa.gov/ssc/data/analysis/documentation/}.}. 
Photon events in a circular region of 15$\degr$ radius centred on the 
position of sources of interest were considered and the {\tt PASS 8} 
instrument response functions (event class 128 and event type 3) 
corresponding to the {\tt P8R2\_SOURCE\_V6} response were used together 
with a zenith angle cut of 90$\degr$. The analysis was performed using 
the {\tt Enrico} Python package \cite{2013arXiv1307.4534S} adapted 
for {\tt PASS 8} analysis. The sky model was constructed based on the 
3FGL catalogue \cite{2015ApJS..218...23A}. The Galactic diffuse 
emission has been modeled using the file {\tt gll\_iem\_v06.fits} 
\cite{2016ApJS..223...26A} and the isotropic background using {\tt 
iso\_P8R2\_SOURCE\_V6\_v06.txt}.
Three energy ranges were considered with the corresponding data cuts in this analysis: 0.1~GeV--500~GeV, 
10~GeV--500~GeV and 50~GeV--500~GeV, with time windows chosen to coincide with 
the \phasetwo observation periods (MJD 56403--56601 and MJD 56441--56513 for \pks and \pg, respectively).

The results of the \fermi spectral analysis, using a simple power-law model, are shown in Fig.~\ref{fig:sed}.
For both AGNs a log-parabola fit did not provide a sufficient improvement to the spectral fit
with respect to the power-law model. Some evidence for a 
softening of the spectrum with energy in the \fermi energy range,
however, was suggested by the analysis of \fermi data for the scan of energy thresholds.
The \fermi analysis results agree very well with the \phasetwo mono data 
within the common overlapping region (80-200~GeV for \pks and 110-200~GeV for \pg).
A strong down-turn spectral feature is apparent in the spectra of both objects at $\approx 100$ GeV, 
i.e. in the transition zone between the two instruments. 

The insets in Fig.~\ref{fig:sed}
provide a comparison of the \phasetwo mono results (shown as blue band) with the
\fermi catalogue data (red for 3FGL \cite{2015ApJS..218...23A}, green for 1FHL \cite{2013ApJS..209...34A}, and purple for 2FHL \cite{2016ApJS..222....5A}).
It is worth noting the considerable agreement between the \fermi contemporaneous 
data (shown on the main plots) and the \fermi catalogue data (shown in the insets), 
particularly for the case of the \pg observations. Since the \fermi 
catalogue data represent the average flux state of the source since data taking 
commenced in 2008, this agreement is suggestive that 
both sources were in average low states of activity during the observational campaign.

\section{EBL deabsorbed spectra}
\label{discussion}

\begin{figure}
  \centering
\includegraphics[width=0.5\linewidth]{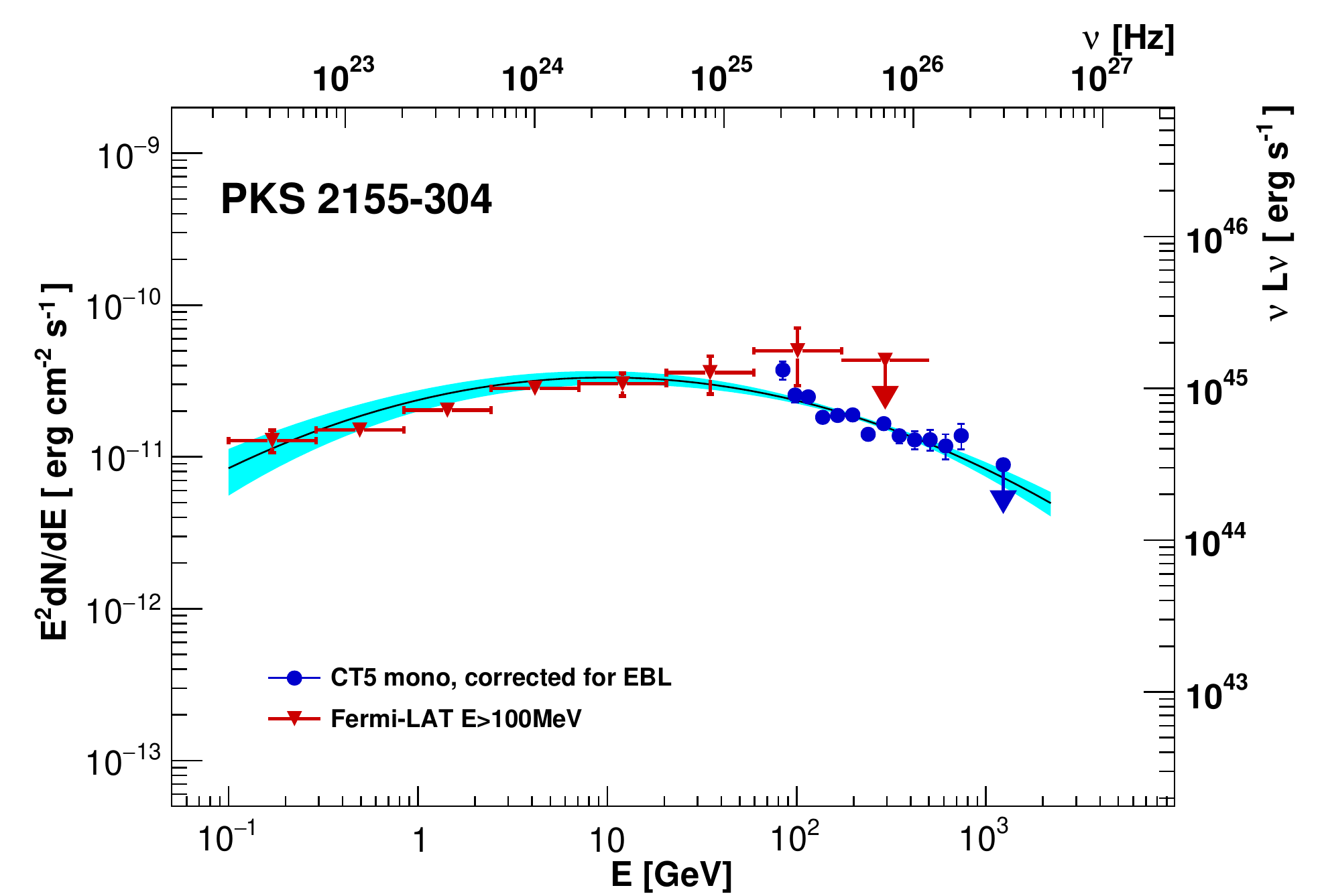}
\includegraphics[width=0.5\linewidth]{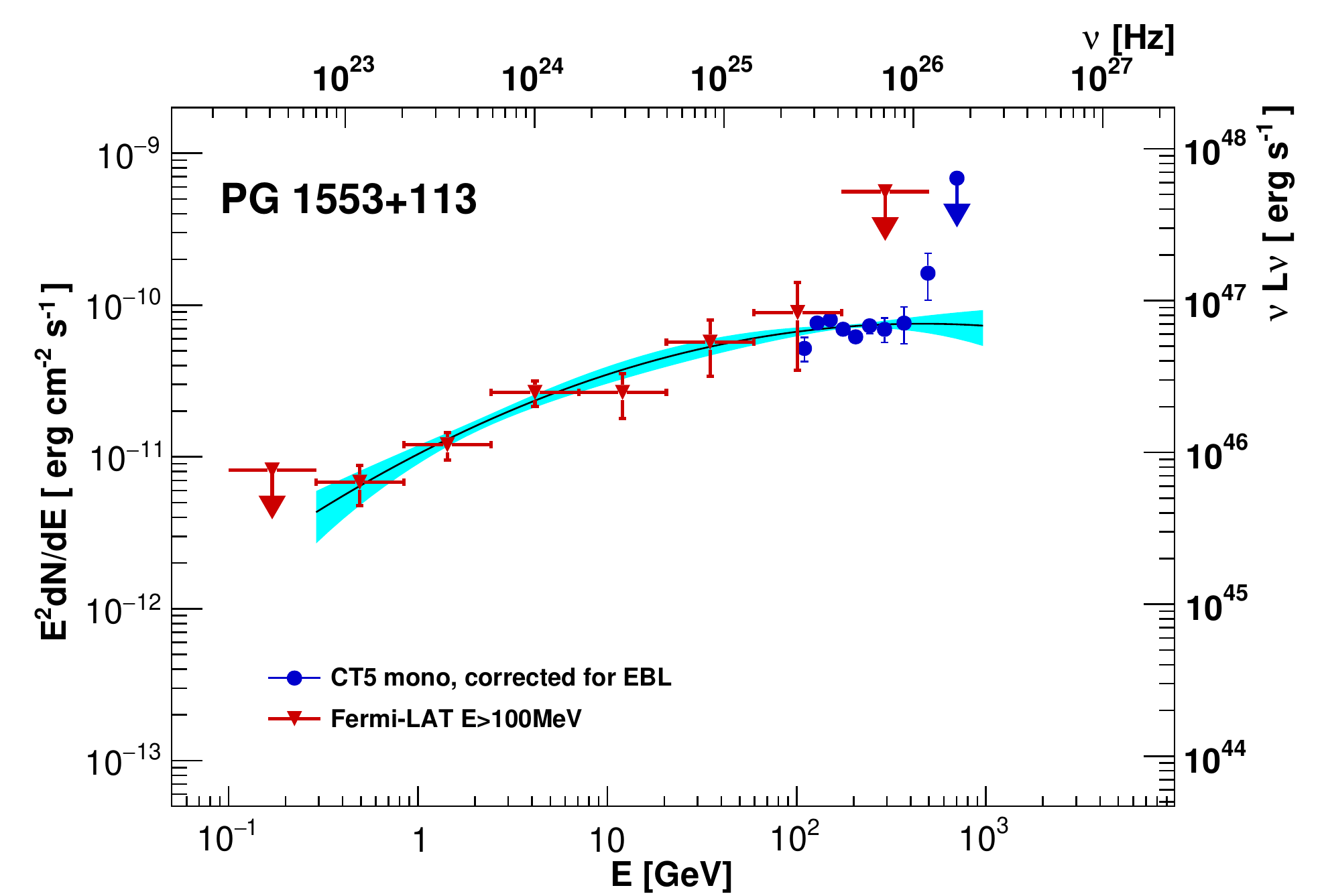}
  \caption{The energy spectrum of \pks (left) and \pg (right) 
obtained from the \phasetwo mono analysis (blue) of the 2013 data corrected for EBL absorption in comparison with the contemporaneous \fermi data with a minimal energy of 0.1~GeV (red). 
The black line is the best-fit log-parabola model to the points and the cyan butterfly indicates 
the 1~$\sigma$ region using only the statistical errors in the combined data set analysis.
The black line is the best-fit log-parabola model to the points and the cyan butterfly indicates the 1~$\sigma$ (statistical error only) uncertainty region.
The right-hand y-axis shows the equivalent isotropic luminosity (not beam corrected).
The assumed redshift of \pg is $z=0.49$. 
}
  \label{fig:sed_ebl_deabsorb}
\end{figure}

The observed spectra carry an imprint of the extragalactic background light,
which leads to suppression of the VHE photon flux due to e+e- pair-production interactions on the way to Earth.
In this work we use the model of Franceschini et al. \cite{2008A&A...487..837F} to calculate the gamma-ray optical depth
so as to reconstruct the intrinsic spectra of the sources.
For this purpose, a spectral model corrected for EBL absorption was fitted simultaneously to the \phasetwo mono and contemporaneous \fermi spectra for both AGNs.
For \pg, whose redshift is not well-constrained, we adopt the well-motivated value of $z = 0.49$ \cite{2015ApJ...802...65A}. 
The resulting spectral fits are shown in Fig.~\ref{fig:sed_ebl_deabsorb}, and the spectral parameters are summarised in Table~\ref{table:Fit}.

In the combined fit procedure, a consideration
of the systematic uncertainties for each of the data sets was included. 
The \phasetwo mono energy scale was found to be the dominant contributing systematic for the deabsorbed spectrum fit results.
The contribution of this uncertainty on the results was estimated through the shifting 
of the data points by an energy scale factor ($\pm$ 19\%) before applying the EBL deabsorption. The variation in 
the best-fit model, introduced via the application of this procedure within the full energy uncertainty
range, was then taken as the systematic uncertainty on each model parameter (see Table~\ref{table:Fit}).
An estimate of the size of the \fermi systematic uncertainties was also obtained, using the 
effective area systematic uncertainty, derived by the LAT collaboration\footnote{see
\url{http://fermi.gsfc.nasa.gov/ssc/data/analysis/scitools/Aeff_Systematics.html}.}. These
uncertainties were noted to be small in comparison to the statistical errors such that their further 
consideration could be safely neglected.

In the case of \pks, separate fits of the \fermi and \phasetwo mono EBL-deabsorbed data 
yield photon indices of $\Gamma=1.82\pm 0.03$ and $\Gamma=2.49\pm 0.05$, respectively.
The power-law model was found to provide a sufficient description in both cases.
The combined fit of the \fermi and \phasetwo mono data was found to prefer a log-parabola model over the power-law model 
at the 5.1~$\sigma$ level. 
The peak flux position within the SED was at moderate energy (around 10~GeV), 
in agreement with its 4-year averaged position found in the 3FGL.

For \pg, an EBL absorbed power-law fit to the \phasetwo mono spectra
required an intrinsic photon index of $\Gamma=1.91\pm 0.13$, 
very close to values suggested by the \fermi spectral fits with thresholds of 100~MeV and 10~GeV.
On the other hand, the fit of the combined \fermi and \phasetwo mono gamma-ray data found a 
log-parabola model preferred at the 2.2~$\sigma$ level over the power-law model
(See Table~\ref{table:Fit} and Figure~\ref{fig:sed_ebl_deabsorb}).
The sizeable systematic errors, once taken into account, however, weaken the preference for the curved model. 
The marginal improvement brought by the log-parabola model suggests that the observed 
softening of the \pg spectrum is predominantly introduced by VHE interaction on 
the EBL, a result consistent with that from other instruments which have searched 
for intrinsic curvature in the source's spectrum \cite{2015MNRAS.450.4399A}. 
Furthermore, the constraint on the intrinsic peak position, at a value of 0.6$^{+1.0}_{-0.4}$~TeV, 
also carries significant uncertainties. 

\begin{table*}
\caption{Parameters obtained for the combined fit of the \fermi and H.E.S.S. data. The reference energy $E_0$ used here is 100 GeV. For both blazars, the log-parabola fits values are provided. For \pg, the values for the power-law model, which was marginally disfavoured, are also given. The last column gives the significance, obtained by comparing the $\chi^2$ values for the log-parabola model against those for the power-law model, using only statistical errors in the analysis.}
\label{table:Fit} 
\centering 
\footnotesize
\begin{tabular}{c c c c c c c} 

\hline
Source &$ \phi_0 [10^{-11}\ {\rm cm}^{-2} {\rm s}^{-1}]$  &  $ \mathbf{\Gamma}$ & $\beta$ &$log_{10}(E_{\rm peak} [{\rm GeV}])$& Sig. ($\sigma$) \\
\hline 
\pks&$ 2.35  \pm 0.10_{\rm stat}\pm 0.57_{\rm sys}$  & $2.30  \pm 0.04_{\rm stat}\pm 0.09_{\rm sys}$ &  $0.15  \pm 0.02_{\rm stat}\pm 0.02_{\rm sys}$ & $0.99 \pm 0.19_{\rm stat}\pm 0.19_{\rm sys}$& 5.1 \\
\pg&$ 5.97\pm 0.25_{\rm stat}\pm 2.19_{\rm sys}$  &  $1.68\pm 0.05_{\rm stat}\pm 0.13_{\rm sys}$ & --  & -- & --\\
\pg&$ 6.66\pm 0.42_{\rm stat}\pm 1.43_{\rm sys}$  &  $1.83\pm 0.08_{\rm stat}\pm 0.29_{\rm sys}$ & $0.12\pm 0.05_{\rm stat}\pm 0.13_{\rm sys}$  & $2.76 \pm 0.45_{\rm stat}\pm 0.93_{\rm sys}$& 2.2 \\

\hline 
\end{tabular}
\end{table*}

\section{Conclusions}
\label{Conclusions}

This work reports on the first study of gamma-ray spectra in extragalactic sources using \phasetwo mono data.
Two bright blazars, \pks and \pg, have been observed in 2013,
resulting in highly confident detections in the monoscopic analysis.
For these results, low-energy thresholds of 80~GeV and 110~GeV, respectively, 
were achieved.
We note that the energy threshold 
remains limited by the accuracy of the background subtraction method,
rather than by the instrument trigger threshold\footnote{This limitation
does not apply to the special case of gamma-ray pulsars,
where the pulsar phasogram can be used to define ``off regions'' for background subtraction.}.
Subsequent improvements and reduction in the energy threshold are therefore considered
possible in the future.

A comparison of the emission level of \pks and \pg with their historic observations 
revealed both to be in low states of activity.
Spectral analysis of the \phasetwo mono data indicates that
a log-parabola fit is statistically preferred over a simple power-law or a broken power-law fit for both AGNs.
The measurement of the curvature parameter for \pg, however, is marginal 
once the systematic errors are taken into account.
Once the \fermi data are also included, 
the presence of a strong spectral downturn feature at an energy of $\sim 100$~GeV is
apparent for both blazars, consistent with previous reports 
\cite{2009ApJ...696L.150A,2014A&A...571A..39H,2010ApJ...708.1310A,2012ApJ...748...46A,2015ApJ...799....7A}.
Such a feature at these energies is expected due to gamma-ray absorption on the EBL during their transit through extragalactic space. 
Correcting for the EBL absorption using the model from \cite{2008A&A...487..837F}
reveals a moderately curved intrinsic spectrum in both cases.
A combined fit of the \fermi and \hess mono data deabsorbed on the EBL
indicates the presence of significant curvature in the intrinsic source spectrum for 
\pks, with the peak of the intrinsic SED sitting at an energy of $\sim 10$~GeV.
A similar EBL deabsorbed analysis for \pg, assuming a redshift of 0.49, reveals a milder level of curvature in the
intrinsic spectrum, suggesting that the peak of the intrinsic SED sits at an energy of $\sim 500$~GeV. 
However, once systematic errors are taken into account, the intrinsic spectrum of \pg was found 
to be consistent with no curvature.
It therefore remains possible that the observed softening in the \pg spectra is purely introduced 
by VHE interaction on the EBL, and is not intrinsic to the source.

Our results demonstrate for the first time the successful 
employment of the monoscopic data from the new \phasetwo instrument (CT5) for blazar studies.
These results mark a significant step forward in lowering the gamma-ray energy range that may be probed with \phasetwo.
This reduction in the energy threshold opens up the opportunity to probe new
low-energy aspects about AGN fluxes, their variability, and their 
attenuation on the EBL out to larger redshifts than that probed 
previously in the \hess~I era.

\section{ACKNOWLEDGMENTS}
{\footnotesize
The support of the Namibian authorities and of the University of Namibia in facilitating the construction and operation of H.E.S.S. is gratefully acknowledged, as is the support by the German Ministry for Education and Research (BMBF), the Max Planck Society, the German Research Foundation (DFG), the French Ministry for Research, the CNRS-IN2P3 and the Astroparticle Interdisciplinary Programme of the CNRS, the U.K. Science and Technology Facilities Council (STFC), the IPNP of the Charles University, the Czech Science Foundation, the Polish Ministry of Science and Higher Education, the South African Department of Science and Technology and National Research Foundation, and by the University of Namibia. We appreciate the excellent work of the technical support staff in Berlin, Durham, Hamburg, Heidelberg, Palaiseau, Paris, Saclay, and in Namibia in the construction and operation of the equipment. This work benefited from services provided by the H.E.S.S. Virtual Organisation, supported by the national resource providers of the EGI Federation.
}

\nocite{*}
\bibliographystyle{aipnum-cp}%
\bibliography{hessII_agn_heidelberg}%

\end{document}